# Magnetic deflection of high-spin sodium dimers formed on helium nanodroplets

*Thomas H. Villers, Benjamin S. Kamerin, Vitaly V. Kresin*

Department of Physics and Astronomy, University of Southern California; Los Angeles, California 90089-0484, USA

**Abstract**

Spectroscopic data on alkali-atom dimers residing on the surface of liquid helium nanodroplets have revealed that they are detected primarily in the weakly bound, metastable, spin-triplet state. Here, by measuring the magnetic Stern-Gerlach deflection of a sodium-doped nanodroplet beam, we transparently demonstrate the abundance of high-magnetic-moment dimers. Their electron spins thermalize with the cryogenic superfluid droplets and become fully oriented by the external magnetic field.

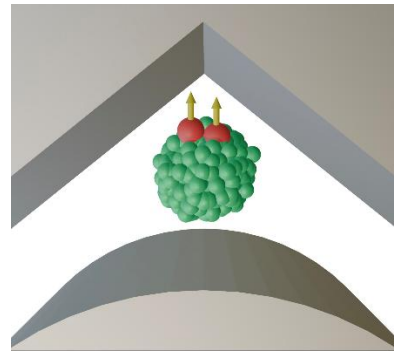

Among key early discoveries in the field of nanodroplet isolation spectroscopy was the observation that alkali dimers and trimers on the surface of superfluid helium nanodroplets are found prevalently in high-spin (triplet or quartet, respectively) states.[1-3] This occurs because alkali atoms aren't wetted by liquid helium[4] and therefore do not sink inside nanodroplets like other impurities, but remain weakly bound in "dimples" on the surface. Hence when such atoms are picked up by the nanodroplet in-flight and encounter each other, the highly exothermic formation of a ground-state dimer or trimer is apt to eject it from the droplet. However, the binding energy of a metastable high-spin dimer or trimer is much lower ($\approx$0.02 eV vs. $\approx$0.7 eV). Therefore the formation of such a configuration deposits less energy into the nanodroplet and consequently is less likely to result in desorption from the surface.[2]

Studies of the formation and dynamics of alkali dimers on superfluid nanodroplets have been summarized in many reviews, for example refs 5-8, and continue to this day.[9-12] Evidence for the formation and dominance of their $S=1$ configuration comes primarily from spectroscopic studies (with additional support provided by measurements of dopant pick-up statistics[13-15]). The evidence is extensive and persuasive, but a measurement supplying a transparently direct demonstration of the high-spin, high-magnetic moment states also can be of value.

Here we present such a measurement. We have performed a magnetic deflection experiment on a beam of sodium-doped helium nanodroplets. The obtained deflection profile immediately and unambiguously puts on view the fact that $Na_2$ molecules residing on nanodroplets have a non-zero magnetic moment and therefore are in a metastable spin-triplet configuration.

The experimental setup is outlined in Fig. 1. Stern-Gerlach deflection of doped helium nanodroplets is performed analogously to electrostatic deflections, previously developed in our group for nanodroplets with polar impurities and described in detail in refs 16-18. Instead of passing between high-voltage electric deflection plates, in the present configuration the beam passes between the poles of a strong deflection magnet.[19] He$_n$ nanodroplets are produced by the expansion of helium gas through a cold nozzle and promptly undergo evaporative cooling to attain an internal temperature of 0.37 K.[20]. They pick up dopants by traversing a heated vapor cell and then enter the deflection field region via a 0.37 mm wide collimating slit. Downstream, the nanodroplets pass another collimating slit and find themselves in the electron-impact ionizer of a quadrupole mass spectrometer. Here the ionized dopants are ejected from their nanodroplets,[21] mass filtered, and detected by an ion counter. By shifting the detector chamber position using a precision motorized stage, we map out the transverse intensity profile of the doped nanodroplet beam. The induced deflection is measured by comparing the beam's "magnet-in" and "magnet-out" profiles.

Some aspects of such a measurement that are specific to an alkali dopant are worth mentioning. A complication derives from the fact that deflections are easier to resolve when the host droplets are less massive, but alkali atoms attach more readily to the surface of larger droplets. Deflections are more pronounced for a slower beam, which calls for a lower nozzle temperature, but colder nozzles generate more massive droplets.[20] By trial and error, the following conditions were adopted: 10 K nozzle temperature and a low helium stagnation pressure of 8 bar (maintained by using a pressure controller), yielding a beam velocity of 255 m/s and a mean nanodroplet size $\langle n \rangle$=6000 atoms. The latter was established by a reference magnetic deflection measurement using FeCl$_2$ doping (ref 18 and unpublished results) of the nanodroplet beam. The distribution of nanodroplet sizes in the beam is taken to be log-normal[20] with a full width of[16] 0.9$\langle n \rangle$.

An important facet of the experiment was that we employed Penning ionization of the alkali impurities, setting the energy of the electron-impact ionizer to 29 eV. At this energy the ion peaks of surface-bound species (such as $Na^+$, $Na_2^+$, $Na_3^+$) formed via interaction with metastable $He^*$, are greatly maximized relative to the appearance of nearby $He_n^+$ background peaks.[22] This allowed us to reduce the resolution of the quadrupole mass filter so as to raise the counting rate of the strongly collimated Na-doped beam while maintaining a high signal-to-background ratio.

The quadrupole mass spectrometer was set to a mass of the $Na_2^+$ ion. In order to ensure that this was not a fragment of a larger cluster, the pick-up cell vapor pressure was gradually raised to a point where the signal of the dimer signal was already visible in the mass spectrum but those of the trimer and larger species were still negligible. For the nanodroplet sizes used here, this corresponded to a sodium vapor pressure of $2\times10^{-4}$ mbar.

The measured magnetic deflection profile is shown in Fig. 2. Straightaway, it serves to establish the main conceptual outcome: the $Na_2$-doped beam is distinctly deflected by a magnetic field, hence the molecule possesses a magnetic moment and is therefore in its high-spin metastable state.

The signal-to-noise ratio of the data is limited by the low intensity (1-10 counts per second) of the strongly collimated beam, due to the reasons outlined above. Nevertheless, a simulation fitting procedure[16-18,23] allows us to extract the absolute value of the magnetic moment, and we find $\mu = 1.9 \pm 0.3$ $\mu_B$ (Bohr magnetons), this is in accord with the value of $2\mu_B$ for the $^3\Sigma$ state of the sodium dimer.

As can be seen from Fig. 2, no significant undeflected component is visible in the in-field profile. Its slight broadening compared to the zero-field profile (full width at half-maximum of 2.5 mm vs. 2.4 mm) is primarily due to the dispersion of nanodroplet sizes in the beam: the smaller, lighter droplets deflect stronger than the larger, heavier ones carrying the same dopant.[16] A two-population fit yielded an estimate of a 5-10 % limit to the presence of ground-state $S=0$ Na dimers. This is compatible with the estimate of 2% for $Na_2$ in the $^1\Sigma$ state in ref 15, but is lower than the 20%-25% fraction of $^1\Sigma$ Rb dimers reported in ref 24. However, the proportion of singlet dimers may depend on the size of the nanodroplet and on the specific alkali metal.

Since only one-sided deflection is observed, rather than three peaks corresponding to $M_S=0,\pm1$, this implies that the magnetic moment is oriented along the applied magnetic field. Hence the dimer is relaxed into its lowest spin component. That is to say, its spin becomes thermalized with the host helium nanodroplet temperature of[20] 0.37 K on a time scale that is much faster than its ≈500 μs travel through the magnetic field. This is consistent with the short, sub-microsecond spin relaxation time found[25,26] for alkali dimers and trimers on nanodroplet surfaces. The mechanism of spin thermalization has not been investigated in detail, but likely proceeds via molecular spin-rotation coupling.[27] Indeed, it is hard to conjecture other pathways for the dissipation of spin Zeeman energy: the helium matrix itself is non-magnetic, and the dimer vibrational level spacings are much larger than this energy. Theoretical guidance toward understanding the rates and pathways of spin relaxation in diatomic and polyatomic molecules on and within superfluid matrices and nanodroplets would be interesting and valuable.

As demonstrated here, the magnetic deflection technique applied to molecules transported by a beam of superfluid helium nanodroplets makes it possible to identify, at a glance as well as quantitatively, the magnetic moment of a cryogenically cold molecule, an important physical observable. In future work, it will be interesting to extend the study of surface-bound species to larger alkali clusters, as well as to heteronuclear alkali dimers and beyond.

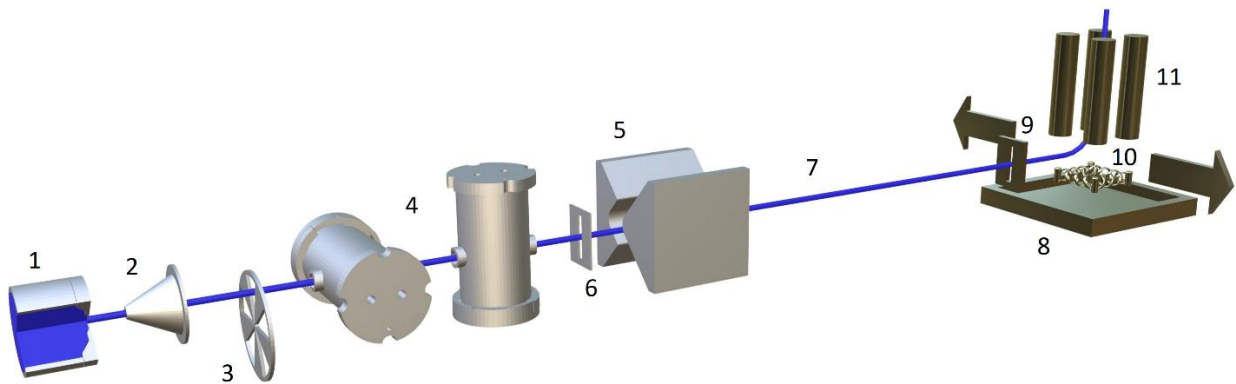

**Figure 1.** An outline of the experimental apparatus (not to scale). 1: Helium expansion nozzle, 2: Skimmer, 3: Beam chopper, 4: Heated pickup cells, 5: Inhomogeneous magnetic field deflector (field 1.1 T, field gradient 330 T/m), 6: Deflector entrance collimator, 7: Free flight path, 8. Detection chamber on a horizontal translation stage, 9: Detector entrance slit, 10: Electron-impact ionizer, 11: Quadrupole mass spectrometer equipped with a pulse-counting channeltron ion detector. The ion pulse counter are synchronized with the beam chopper, enabling reliable signal-to-background discrimination even for dopant ion signals of several counts per second. The chopper is also used to perform time-of-flight measurement of the beam velocity distribution. Conveniently for deflection measurements, since the nanodroplet beam is formed in a supersonic expansion, its velocity distribution is very narrow.

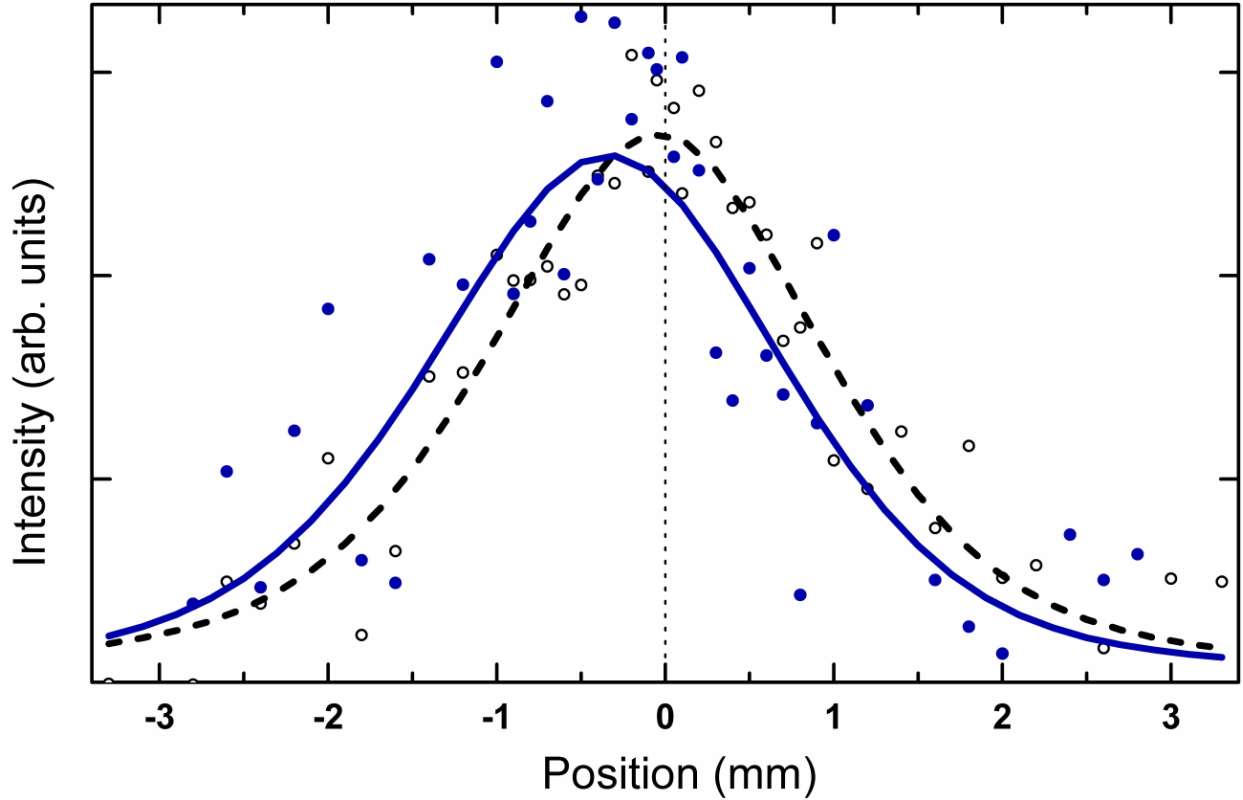

**Figure 2.** Stern–Gerlach deflection profiles of helium nanodroplets doped with Na$_2$. The open circles and the dashed line are the zero-field beam profile data (see the main text) and their smoothing fit, respectively, centered on the beam axis denoted by the dotted line. The solid circles are the deflected beam profile data. The solid line is a simulation fit of the deflection process which yields the molecule's magnetic moment. The two sets of data have been scaled to the same integrated intensity; we have verified that this has minimal effect on the magnitude of the fitted moment.

# References

(1) Stienkemeier, F.; Ernst, W. E.; Higgins, J.; Scoles, G. On the use of Liquid Helium Cluster Beams for the Preparation and Spectroscopy of the Triplet States of Alkali Dimers and Other Weakly Bound Complexes. *J. Chem. Phys.* **1995,** 615-617.

(2) Higgins, J.; Ernst, W.E.; Callegari, C.; Reho, J.; Stienkemeier, F.; Lehmann, K.K.; Scoles, G.; Gutowski, M. Spin Polarized Alkali Clusters: Observation of Quartet States of the Sodium Trimer. *Phys. Rev. Lett.* **1996**, *77*, 4532-4535.

(3) Higgins, J.; Callegari, C.; Reho, J.; Stienkemeier, F.; Ernst, W.E.; Lehmann, K.K.; Gutowski, M.; Scoles, G. Photoinduced Chemical Dynamics of High-Spin Alkali Trimers. *Science* **1996**, *273*, 629-631.

(4) Ancilotto, F.; Cheng, E.; Cole, M. W.; Toigo, F. The Binding of Alkali Atoms to the Surfaces of Liquid Helium and Hydrogen. *Z. Phys. B* **1995**, *98*, 323-329.

(5) Tiggesbäumker, J.; Stienkemeier, F. Formation and Properties of Metal Clusters Isolated in Helium Droplets. *Phys. Chem. Chem. Phys.* **2007**, *9*, 4748-4770.

(6) Callegari, C.; Ernst, W. E. Helium Droplets as Nanocryostats for Molecular Spectroscopy– from the Vacuum Ultraviolet to the Microwave Regime. In: *Handbook of High Resolution Spectroscopy;* Merkt, F., Quack, M., Eds.; Wiley: Chichester, 2011; Vol. 3, pp. 1551-1594.

(7) Mudrich, M.; Stienkemeier, F. Photoionisaton of Pure and Doped Helium Nanodroplets. *Int. Rev. Phys. Chem.* **2014**, *33*, 301–339.

(8) Bruder, L.; Koch, M.; Mudrich, M.; Stienkemeier, F. Ultrafast Dynamics in Helium Droplets. In: *Superfluid Helium Nanodroplets: Spectroscopy, Structure, and Dynamics*; Slenczka, A., Toennies J. P., Eds.; Springer: Cham, 2022, pp. 447-511.

(9) Albertini, S.; Martini, P.; Schiller, A.; Schöbel, H.; Ghavidel, E.; Oncák, M.; Echt, O.; Scheier, P. Electronic Transitions in $Rb_2^+$ Dimers Solvated in Helium. *Theor. Chem. Acc*. **2021**, *140*, 29.

(10) Kristensen, H. H.; Kranabetter, L.; Schouder, C. A.; Stapper, C.; Arlt, J.; Mudrich, M.; Stapelfeldt, H. Quantum-State-Sensitive Detection of Alkali Dimers on Helium Nanodroplets by Laser-Induced Coulomb Explosion. *Phys. Rev. Lett.* **2022**, *128*, 093201.

(11) Kranabetter, L.; Kristensen, H. H.; Ghazaryan, A.; Schouder, C. A.; Chatterley, A. S.; Janssen, P.; Jensen, F.; Zillich, R.E.; Lemeshko, M.; Stapelfeldt, H. Nonadiabatic Laser-Induced Alignment Dynamics of Molecules on a Surface. *Phys. Rev. Lett*. **2023**, *131*, 053201.

(12) Albrechtsen, S. H.; Christensen, J. K.; Tanyag, R. M. P.; Kristensen, H. H.; Stapelfeldt, H. Laser-Induced Coulomb Explosion of Heteronuclear Alkali-Metal Dimers on Helium Nanodroplets. *Phys. Rev. A* **2024**, *109*, 043112.

(13) Vongehr, S.; Kresin, V. V. Unusual Pick-up Statistics of High-Spin Alkali Agglomerates on Helium Nanodroplets. *J. Chem. Phys*. **2003**, *119*, 11124–11129.

(14) Nagl, J.; Auböck, G.; Hauser, A. W.; Allard, O.; Callegari, C.; Ernst, W. E. High-Spin Alkali Trimers on Helium Nanodroplets: Spectral Separation and Analysis. *J. Chem. Phys*. **2008**, *128*, 154320.

(15) Bünermann, O.; Stienkemeier, F. Modeling the Formation of Alkali Clusters Attached to Helium Nanodroplets and the Abundance of High-Spin States. *Eur. Phys. J. D* **2011**, *61*, 645–655.

(16) Niman, J. W.; Kamerin, B. S.; Merthe, D. J.; Kranabetter, L.; Kresin, V. V. Oriented Polar Molecules Trapped in Cold Helium Nanodroplets: Electrostatic Deflection, Size Separation, and Charge Migration. *Phys. Rev. Lett*. **2019**, *123*, 043203.

(17) J. W. Niman, PhD thesis, University of Southern California (2022).

(18) B. S. Kamerin, PhD thesis, University of Southern California (2023).

(19) Liang, J.; Fuchs, T. M.; Schäfer, R; Kresin, V. V. Strong Permanent Magnet Gradient Deflector for Stern–Gerlach-Type Experiments on Molecular Beams. *Rev. Sci. Instrum*. **2020**, *91*, 053202.

(20) Toennies, J. P. Helium Nanodroplets: Formation, Physical Properties and Superfluidity. In: *Superfluid Helium Nanodroplets: Spectroscopy, Structure, and Dynamics*; Slenczka, A., Toennies J. P., Eds.; Springer: Cham, 2022, pp. 1-40.

(21) A. Mauracher, O. Echt, A. M. Ellis, S. Yang, D. K. Bohme, J. Postler, A. Kaiser, S. Denifl, and P. Scheier, Cold physics and chemistry: Collisions, ionization and reactions inside helium nanodroplets close to zero K, Phys. Rep. **751**, 1 (2018).

(22) Vongehr, S.; Scheidemann, A.A.; Wittig, C.; Kresin, V. V. Growing Ultracold Sodium Clusters by Using Helium Nanodroplets. *Chem. Phys. Lett*. **2002**, *353*, 89–94.

(23) The strength of the deflected profile was approximately 20% lower than the undeflected one. This behavior is possibly connected with the atypically low helium gas stagnation pressure that needed to be employed in this measurement, and may reflect either a decay in the beam intensity or the presence of a bimodal nanodroplet size distribution. However, we have verified that the best-fit magnetic moment value of the deflected $Na_2$ remains the same for simulations using either the experimentally measured intensity or the one rescaled to initial strength.

(24) Kristensen, H. H.; Kranabetter, L.; Schouder, C. A.; Stapper, C.; Arlt, J.; Mudrich, M.; Stapelfeldt, H. Quantum-State-Sensitive Detection of Alkali Dimers on Helium Nanodroplets by Laser-Induced Coulomb Explosion. *Phys. Rev. Lett*. **2022**, *128*, 093201.

(25) Nagl, J.; Auböck, G.; Callegari, C.; Ernst, W. E. Magnetic Dichroism of Potassium Atoms on the Surface of Helium Nanodroplets. *Phys. Rev. Lett.* **2007**, *98***,** 075301.

(26) Koch, M.; Auböck, G.; Callegari, C.; Ernst, W. E. Coherent Spin Manipulation and ESR on Superfluid Helium Nanodroplets. *Phys. Rev. Lett.* **2009**, *103*, 035302.

(27) Fuchs, T. M.; Schäfer, R. Effect of Vibrational Excitation and Spin-Rotation Coupling on Stern-Gerlach Experiments: A Detailed Case Study on $GdSn_{15}$ as an Asymmetric Rotor. *Phys. Rev. A* **2019**, *100*, 012512.